\documentclass[aps,prb,twocolumn,showpacs,amsmath,amssymb,]{revtex4-2}
\usepackage{graphicx}

\usepackage{hyperref}
\hypersetup{  colorlinks=true,     urlcolor=blue,     citecolor=blue,     linkcolor=blue,}

\def\NO{(TMTTF)$_2$NO$_3$}
\def\cm{${\rm cm}^{-1}$}

\begin{document}
\title{Charge- and anion-orderings in the
quasi-one-dimensional organic conductor (TMTTF)$_2$NO$_3$}
\date{\today}

\author{Olga~Iakutkina}
\affiliation{1.~Physikalisches Institut, Universit{\"a}t Stuttgart, 70569 Stuttgart, Germany}
\author{Lena Nadine~Majer}
\affiliation{1.~Physikalisches Institut, Universit{\"a}t Stuttgart, 70569 Stuttgart, Germany}
\author{Guilherme Gorgen Lesseux}
\affiliation{1.~Physikalisches Institut, Universit{\"a}t Stuttgart, 70569 Stuttgart, Germany}
\author{Gabriele Untereiner}
\affiliation{1.~Physikalisches Institut, Universit{\"a}t Stuttgart, 70569 Stuttgart, Germany}
\author{Martin Dressel}
\affiliation{1.~Physikalisches Institut, Universit{\"a}t Stuttgart, 70569 Stuttgart, Germany}

\begin{abstract}
The quasi-one-dimensional organic conductors (TMTTF)$_2X$ with non-centrosymmetric anions commonly undergo charge- and anion-order transitions upon cooling. While for compounds with tetrahedral anions ($X$ = BF$_4^-$, ReO$_4^-$, and ClO$_4^-$) the charge-ordered phase is rather well understood, the situation is less clear in the case of planar triangular anions, such as (TMTTF)$_2$NO$_3$. Here we explore the electronic and structural transitions by transport experiments, optical and magnetic spectroscopy. This way we analyze the temperature dependence of the charge imbalance 2$\delta$ and an activated behavior of $\rho(T)$
with $\Delta_{\rm CO}\approx 530$~K below $T_{\rm CO} = 250$~K.
Since (TMTTF)$_2$NO$_3$ follows the universal relation between charge imbalance 2$\delta$
and size of the gap $\Delta_{\rm CO}$, our findings suggest that charge order is determined by TMTTF stacks with little influence of the anions.
Clear signatures of anion ordering are detected at $T_{\rm AO}=50$~K. The tetramerization affects the dc transport, the vibrational features of donors and acceptors, and leads to formation of spin singlets.
\end{abstract}.


\maketitle

\section{\label{sec:introduction}Introduction}
In low-dimensional electron systems the influence of electronic correlations becomes more pronounced
and may govern the effects of the crystal lattice \cite{GiamarchiBook,BaeriswylDegiorgi04}.
A prime example is given by the quasi-one-dimensional Fabre salts (TMTTF)$_2X$, where TMTTF
stands for tetramethyltetratiofulvalene and $X$ is a monovalent anion.
By stoichiometry these systems possess three-quarter-filled conduction bands but dimerization leads to effectively half-filling. Although they are supposed to be metals according to their band structure,
electron-electron interaction leads to charge localization.
Furthermore, different broken-symmetry ground states can be realized depending on the subtle interplay of charge, spin, and lattice degrees of freedom \cite{jerome1982organic,FargesBook, dressel2003spin,*dressel2007ordering, TomicDressel15,pouget2018donor,*DresselTomic2020}.

In (TMTTF)$_2X$ salts with 
octahedral anions ($X$ = PF$_6^-$, AsF$_6^-$, SbF$_6^-$, and TaF$_6^-$) charge order (CO) was extensively investigated by several methods \cite{chow2000charge,*zamborszky2002competition,*yu2004electron,Monceau2001,*Nad2006,*Monceau2012, dumm2004magnetic,*dumm2005mid,*dressel2012comprehensive,oka2015charge, nakamura2003possible,*Furukawa2005,*Nogami2005,*Iwase2011,Salameh2011,*Yasin2012,*Dressel2012ESR}.
An regular 1010 charge pattern develops along the molecular stacking axis $a$ below the transition temperature $T_{\rm CO}$. For non-centrosymmetric anions ($X$ = BF$_4^-$, ReO$_4^-$, and ClO$_4^-$) situation is more complicated because --~in addition to the CO phase transition~-- the
anion may order (AO) in an alternating fashion at low temperatures, which leads to a tetramerization of the donor molecules \cite{coulon1982new,moret1983structural,pouget1982x,*ravy1986structural,rosslhuber2018structural} and doubling of the charge pattern to alternating sequence 0110 below $T_{\rm AO}$. Despite the different anion size, for all these compounds a universal relation between charge imbalance 2$\delta$, ordering temperature $T_{\rm CO}$ and charge gap $\Delta_{\rm CO}$ was revealed  \cite{pustogow2016electronic}, which implies that the CO state is intrinsic for the TMTTF stacks and mainly determined by a competition between the inter-site Coulomb repulsion $V$ and the bandwidth $W$.

So far, the detailed investigation were confined to (TMTTF)$_2X$ salts with tetrahedral anions, while other symmetries were considered only incidentally.
For (TMTTF)$_2$\-SCN with linear anions, CO coincides with the ordering of the non-centrosymmetric anions \cite{Yasin2012}. The situation is even more complex in the case of $X$ = NO$_3^-$, where   the planar anions have trigonal symmetry.
For decades it is discussed in the community whether or not charge order occurs in \NO\
\cite{coulon1982new,coulon2015electronic}. Like other salts with non-centrosymmetric anions, the salt undergoes a AO transition around 50~K, which was identified by X-ray diffraction and ESR measurements \cite{pouget1982x,coulon2015electronic}. Recent ESR data provided further evidence for the stabilization of a CO state below 250~K \cite{majer2020charge}.
Here we present the results of detailed spectroscopic investigations of the charge distribution in \NO\ in the CO and AO states.

\begin{figure}[h]
\centering
\includegraphics[width=\columnwidth]{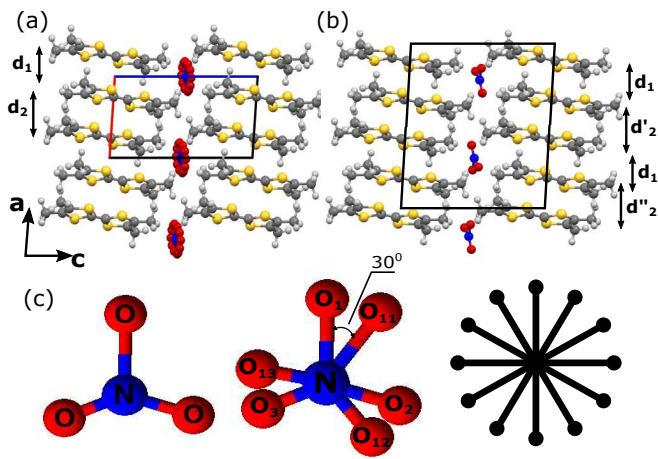}
\caption{(a) Crystal structure of \NO\ where TMTTF stands for tetramethyltetrathiafulvalene at the normal state with finite dimerization and disordered anions, $d_1$ and $d_2$ are intra- and inter- dimmers distances respectively. (b)~Suggested low-temperature state with alternating orientation of the anions that leads to a doubling of the unit cell and tetramerization along the TMTTF molecular stack. (c)~From the left to the right shape of the planar NO$_3^-$ anion, the difference between two anion orientations, and all possible orientations are shown.}
\label{fig:structure}
\end{figure}

\section{Materials and Experiments}

The aciculate single crystals of \NO\ used for our investigations were grown by standard electrochemical methods \cite{kohler2011comprehensive}. The organic tetramethyltetratiofulvalene (TMTTF) donor molecules are stacked along the crystallographic $a$-axis;
these stacks are separated by monovalent NO$_3^-$ anions in the $c$-direction, as displayed in Fig.~\ref{fig:structure}(a). In the normal state at ambient condition, the molecules are slightly dimerized along the chains with intra- and inter- dimer distances $d_1$ and $d_2$ respectively.

At elevated temperature, the planar NO$_3^-$ anions exhibit orientational disorder with four possible arrangements for the nitrate ions, with each possible orientation rotated on 30$^{\circ}$ with respect to each other, as indicated in Fig.~\ref{fig:structure}(c). Since there are in total twelve possible positions for oxygen, the anion almost resembles  a nitrogen atom surrounded by an oxygen ring \cite{liautard1982etude}.
At reduced temperatures, below $T_{\rm AO}$, the anions arrange in an orderly fashion. For the metallic analogue (TMTSF)2NO3,
which does not exhibit any ordering down to T$_{AO}$ = 45 K,
the crystal structure in the anion-ordered state was resolved: here the anion order leads to a doubling of the
unit cell \cite{barrans1999low,Guster_2020}. For (TMTTF)2NO3 we expect that the
anion ordering leads to tetramerization of the TMTTF
molecules with distances $d_1$, $d_2^{\prime}$, $d_1$, $d_2^{\prime\prime}$ as shown in Fig.~\ref{fig:structure}(b).

In order to characterize the compound by dc resistivity measurements, two contacts were attached along the $a$-direction by carbon paste. To ensure good thermal contact, the samples were anchored to a sapphire plate and slowly cooled down to $T=20$~K. The voltage applied to the crystal was in the range of 0.5 to 5~V;
from the measured current, the dc resistivity is calculated as a function of temperature, $\rho(T)$.

\begin{figure}[b]
\centering
\includegraphics[width=0.9\columnwidth,clip]{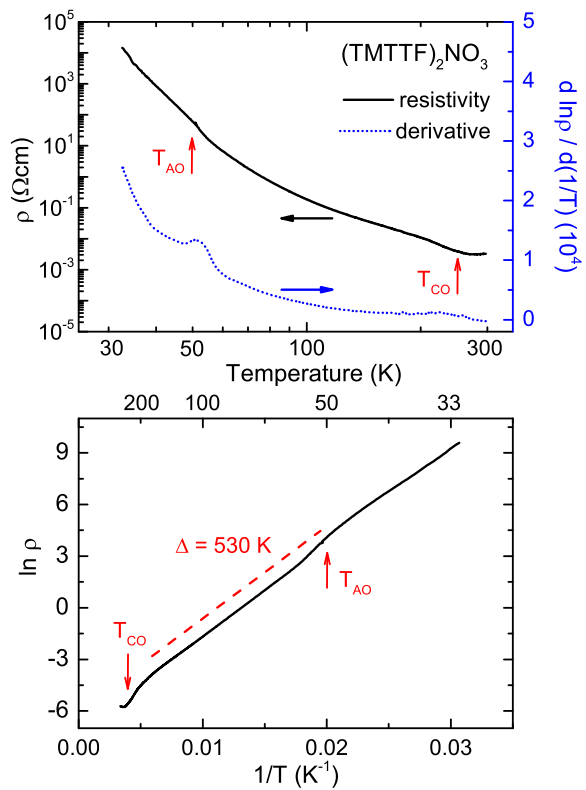}
\caption{Transport properties of \NO\ along the chain direction $a$. (a)~The black solid curve corresponds to the temperature dependence of the dc resistivity; note the double logarithmic scale on the left and bottom axis.
The blue dotted curve gives the derivative d\,ln$\rho$/d$({1}/{T})$ corresponding to the linear right axis.
(b)~The Arrhenius plot of the temperature-dependent resistivity allows the extraction of an activation energy $\Delta = 530$~K, indicated by the dashed red line.}
\label{fig:transport}
\end{figure}

To learn about the charge disproportionation in \NO, optical spectroscopy was performed
utilizing a Bruker Hyperion infrared microscope attached to a Bruker Vertex 80v Fourier-transform infrared spectrometer.
Reflectivity spectra $R(\omega,T)$ were collected in a frequency range from 500 to 8000~\cm\ between room temperature and 10~K. The optical conductivity $\sigma(\omega)$ was calculated using Kramers-Kronig analysis with constant $R(\omega)$ extrapolation below 500~\cm\ and $\omega^{-4}$ decay for higher frequencies. We first  focused on the charge-sensitive infrared-active intramolecular vibrational mode
$\nu_{28}({\rm B}_{1u})$ \cite{meneghetti1984vibrational}. These measurements were performed with the light polarized along the $c$-direction, i.e.\ perpendicular the main plane. As the amount of charge per TMTTF molecule is
linearly related to the resonance frequency of the $\nu_{28}$ mode, charge imbalance in CO and AO states can be directly probed by tracking the peak positions upon cooling \cite{dressel2012comprehensive}.
In addition, we performed reflection measurements off the crystal plane with the light polarized along the stacking direction (i.e. $E\parallel a$).

We complemented our investigations by measurements of the electron spin resonance (ESR) down to low temperatures. The angular-dependence of the $g$-factor and linewidth $\Delta H$ was obtain by a  X-band  spectrometer (Bruker  EMXplus) as presented in detail in Ref.~\onlinecite{majer2020charge}.

\section{Results and Analysis}

\subsection{Transport properties}

The solid black curve in Fig.~\ref{fig:transport}(a) represents the tem\-per\-a\-ture-dependent dc resistivity of \NO\ measured along the $a$-axis; the overall behavior is reproduced on several single crystals. It is plotted in a double-log\-a\-rith\-mic fashion to display the overall behavior.
To identify the phase transitions more precisely, we calculate the derivative ${\rm d}\ln\rho/{\rm d}({1}/{T})$ and plot it as a function of temperature by the dotted blue curve \cite{rosslhuber2018structural}.
A most pronounced peak in ${\rm d}\ln\rho/{\rm d}({1}/{T})$ is found at the AO transition around $T_{\rm AO}= 50$~K. The absence of any sign of hysteresis resembles more a second-order instead of a structure transition of first order.
The broad minimum of $\rho(T)$ at elevated temperature corresponds to the crossover from metal-like behavior to a charge-localized state; the zero-crossing in the derivative occurs at $T_{\rho}\approx 280$~K, which is slightly higher than what was first reported by Coulon et al. \cite{coulon1982new}
The CO transition temperature $T_{\rm CO} \approx 250$~K is also best identified in the local maximum in ${\rm d}\ln\rho/{\rm d}({1}/{T})$, which occurs when the resistivity exhibits a rapid increase around 225~K.
Note, in \NO\ salt the CO transition is more gradual compare to other Fabre salts because with decreasing anion size the interchain separation becomes smaller, and thus one-dimensionality is less pronounced \cite{knoblauch2012charge,oka2015charge}.
In Fig.~\ref{fig:transport}(b) the Arrhenius plot of the resistivity is displayed, i.e.\
$\ln\rho(T)$ versus inverse temperature $1/T$. The CO gap $\Delta_{\rm CO}$ can be extracted.
The activated behavior continues with the same activation energy $\Delta_{\rm CO} \approx 530$~K below the AO transition.

\subsection{Magnetic properties}
\begin{figure}[h]
\centering
\includegraphics[width=0.9\columnwidth,clip]{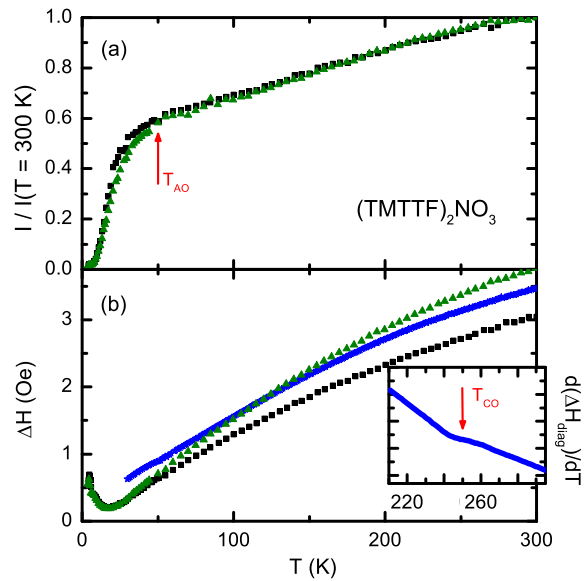}
\caption{ESR parameters of \NO\ obtained from X-band measurements as a function of temperature.
(a) The intensity $I(T)$ for the $a$- and $b$-orientation normalized to the room temperature value. The ordering of the anions at $T_{\rm AO}= 50$~K causes a drop of the spin susceptibility.
(b)~The temperature dependent linewidth $\Delta H$ meausred along different directions:
$a$-axis (black squares), $b$-direction (green triangles), and diagonal direction (blue stars).
Strong evidence for
the presence of charge order in this compound comes from
the fact, that at T$_{CO}$ = 250 K the slope in the diagonal
direction changes, as clearly shown in the inset, where the
derivative of the smoothed data is plotted. This leads to a
crossing (at around 125 K) of the temperature dependence
for the linewidth obtained in the diagonal direction and the
linewidth for the b-orientation. }
\label{fig:ESR2}
\end{figure}

From our temperature dependent X-band ESR experiments we extract the intensity, resonance frequency and the line width \cite{majer2020charge}. Fig.~\ref{fig:ESR2}(a) displays the ESR intensity $I(T)$ normalized to the room temperature value $I_{300 {\rm K}}$, which corresponds to the spin susceptibility $\chi_S(T)$.
The overall behaviour resembles the spin susceptibility of other (TMTTF)$_2X$ compounds \cite{Dumm2000a,*Salameh2011,*Yasin2012}. The intensity
monotonically decreases upon cooling from room temperature to the anion-ordering transition at T$_{AO} = 50$ K,
below which the chain dimerizes accompanied by the formation of spin singlets, leading to the observed exponential decay of $\chi_{S}$(T) \cite{moret1983structural,coulon2015electronic,Dumm2000a,Dumm2000b}. The extracted spin
gap $\Delta_{\sigma} \approx$ 48 K is similar for all three crystallographic
directions and in agreement with previous reports \cite{coulon2015electronic}.

At a first glance, the temperature dependence of the linewidth $\Delta H(T)$ plotted in Fig.~\ref{fig:ESR2}(b) exhibits no distinct features along any crystallographic axes; in fact, the behavior resembles the one known from (TMTTF)$_2$PF$_6$ where also no kink become obvious at $T_{\rm CO}$ \cite{Yasin2012}. A closer look, however, reveals that for the diagonal orientation a change in the slope for the derivative [d($\Delta H_\text{diag}$)/d$T$] occurs
at $T_{\rm CO}=250$~K, as indicated in the inset of Fig.~\ref{fig:ESR2}.
This observation is supported by the angular dependence of the ESR data reported in \cite{majer2020charge}.

Here we plot the linewidth measured along the $a$-axis, the $b$-orientation as well as along $45^{\circ}$, i.e.\ between the $a$- and $b$-direction. Already around $T = 250$~K the line of
the diagonal-direction is broader than expected. At $T \approx 125$~K the
linewidth in the diagonal direction of the $ab$-plane becomes larger than along the
$a$- and $b$-directions. This is taken as indication that a charge-order transition is
present in \NO\ .

\subsection{Vibrational spectroscopy}

\begin{figure*}[t]
\centering
\includegraphics[width=0.8\textwidth, clip]{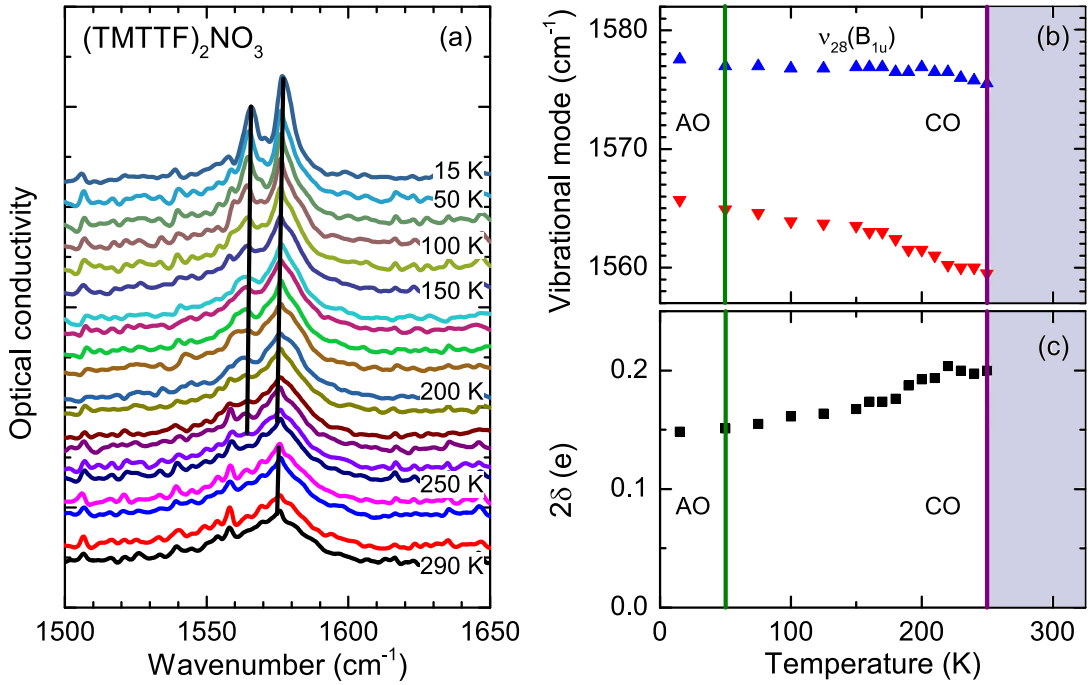}
\caption{(a)~Temperature evolution of the charge-sensitive infrared-active molecular vibration $\nu_{28}$ observed in the mid-infrared spectra of \NO\ for $E\parallel c$. For clarity reasons, the curves are shifted with respect to each other.
(b)~Resonance frequency of the $\nu_{28}({\rm B}_{1u})$ mode as a function of temperature obtained from fitting the spectra with two Lorentzians below T$_{\rm CO}$ (blue and red triangles). The peaks gradually diminish upon raising the temperature and cannot be reliably distinguished
above T$_{CO}$ = 250 K. The grey area indicates that we present the fit results only below 250 K. The overall behavior resembles a
gradual phase transition more than a rapid jump common to first-order transition. Note the slight upwards shift for $T<T_{\rm AO}$.
(c)~By using Eq.~(\ref{eq:nu28}) the charge imbalance $2\delta$ can be directly calculated from the peak separation $\Delta\nu$.}
\label{fig:v28}
\end{figure*}

Out of the molecular vibrations in TMTTF, three modes are most sensitive to the charge per molecule.
Two of them are totally symmetric A$_g$ ($\nu_3$ and $\nu_4$), and as a result they are infrared inactive; however they become visible via electron-molecular vibrational (emv) coupling \cite{dressel2012comprehensive}.
The third one is an asymmetric infrared active B$_{1u}$ ($\nu_{28}$) mode which can be probed in out-of-plane direction (i.e.\ $E \parallel c$-axis). It exhibits a linear shift of the resonance frequency with ionicity of the TMTTF molecule due to the strengthening/loosening of the intramolecular bonds.
The rather large intensity and weak coupling to the electronic background makes it ideal for investigation of the charge distribution. In the normal state, each TMTTF molecule contains the net charge $\rho=+0.5e$, which corresponds to one
peak in the infrared spectra ($\nu^{+0.5} = 1547$~\cm).

\begin{figure}[h]
\centering
\includegraphics[width=0.9\columnwidth,clip]
{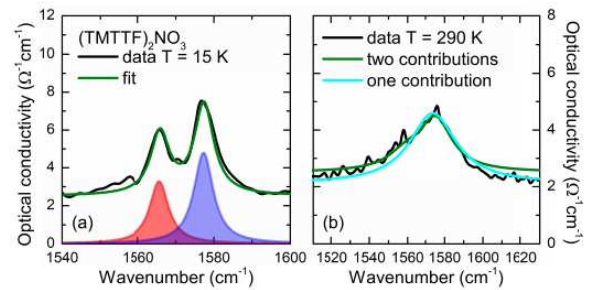}
\caption{Vibrational spectra of \NO\ in the range
of the $\nu_{28}$ mode. (a) The 15 K spectrum can be readily fitted
with two Lorentzians, indicated by the red and blue curves;
the overall fit is represented by the green line. (b) At T =
290 K the curve is much broader (note, the frequency scale
differs by a factor of 2) and the mode less pronounced. We
present a fit of the data with one contribution (cyan line)
and two contributions (green line) and conclude that a two-component fit is not sensible.}
\label{fig:fitting}
\end{figure}

In Fig.~\ref{fig:v28}(a) the development of the $\nu_{28}$(B$_{1u}$) mode is shown as the \NO\ crystal is cooled down. While at room temperature we observe only one broad band that cannot be sensibly separated, two peaks are clearly distinguished below $T=250$~K. The $\nu_{28}$ mode in \NO\ is much broader compare to corresponding features observed in compounds with octahedral anions \cite{dressel2012comprehensive}, and comparable to those seen for compounds with tetrahedral anions \cite{pustogow2016electronic}. This can be attributed to the inherent disorder of the anions.
To  analyze the data qualitatively, we fit the $\nu_{28}$ mode using two Lorentz oscillators, which works reasonably well up to $T_{\rm CO}$. In Fig.~\ref{fig:v28}(b) the resonance frequencies are depicted as a function of temperature. Due to the increasing line width and
diminishing intensity, it was not possible to reliably separate two contributions above 250 K, as illustrated in Fig.\ref{fig:fitting}. The absence of a
sensible fit results is indicated by the shaded area. Note,
that this is a gradual process resembling a continuous
second-order phase transition. In addition, we can identify a simultaneous upshift of both modes by less than 1~\cm\ as the AO state is entered below 50~K; also the intensity of the modes increases. From the positions of the two peaks, the charge imbalance $2\delta$ can be calculated via
\begin{equation}
2\delta = \frac{\Delta\nu}{80~{\rm cm}^{-1}} \quad ,
\label{eq:nu28}
\end{equation}

where $\Delta\nu$ is the spectral separation between the peaks in the CO state.

The temperature dependence of $2\delta(T)$ is plotted as a function of temperature in Fig.~\ref{fig:v28}(c). It becomes obvious at $T_{\rm CO}=250$~K and reaches the maximum $2\delta=(0.21\pm 0.01)e$ at 220~K; this perfectly matches the temperatures obtained from the transport data [Fig.~\ref{fig:transport}(a)].
The charge disproportionation then gradually decreases to $0.15e$, and remains unchanged in the AO state.

\section{Discussion}

\subsection{Charge order transtition}
In the case of Fabre salts with highly symmetric anions, such as (TMTTF)$_2$SbF$_6$ and (TMTTF)$_2$AsF$_6$,
a noticeable kink in the ESR linewidth has been identified at $T_{\rm CO}$ \cite{Yasin2012}.
The charge disproportionation on the TMTTF molecules leads to nonequivalent couplings between the anions and the TMTTF molecules that influences the magnetic behavior of the spin chain.
The ESR linewidth of \NO, however, barely changes at $T_{\rm CO}$; only a plateau in the derivative is identified in the inset of Fig.~\ref{fig:ESR2}.

The low-temperatures optical spectra of \NO\ displayed in Fig.~\ref{fig:v28}(a) contain two
vibrational features corresponding to $\nu_{28}$ that reflect the presence of two TMTTF sites containing non-equivalent charge; the strength of the low-frequency feature gradually diminishes upon warming up, can barely be resolved above 200~K and becomes undistinguishable around $T_{\rm CO}$.
Traces of charge disproportionation might be already present at room temperature, as observed in some BEDT-TTF compounds~\cite{Ivek2011,Pustogow2019a,*Pustogow2019b}, for instance; however, the width of the vibrational band and the weakness of the second feature does not allow a firm conclusion.
Beyond this uncertainty, $2\delta(T)$ continuously decreases when cooling below $T_{\rm CO}$ before it saturates around 0.15$e$, as depicted in Fig. \ref{fig:v28}(c).
It stays almost the same in the AO state, and we can conclude that the effect of the AO on charge distribution is rather small.

\begin{figure}
\centering
\includegraphics[width=0.8 \columnwidth,clip]{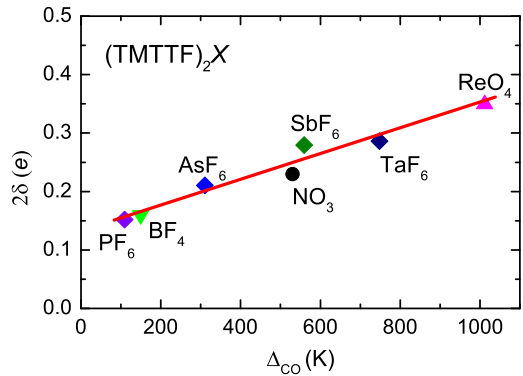}
\caption{Amount of charge disproportionation $2\delta$ plotted versus the charge-order gap $\Delta_{\rm CO}$ obtained from temperature-dependent transport measurements \cite{chow2000charge,oka2015charge}. The red line indicates the linear dependence between $2\delta$ and $\Delta_{\rm CO}$.}
\label{fig:2delta}
\end{figure}

From a comparative study \cite{pustogow2016electronic} Pustogow {\it et al.} concluded that the CO state in (TMTTF)$_2X$ salts with octahedral and tetrahedral anions is determined by the stacking of the organic molecules
and independent on symmetry of the anions. In other words, the CO state is only governed by the competition between intersite Coulomb repulsion $V$ on the one hand and overlap integral $t$ or bandwidth $W$ on the other hand. In first approximation, the charge imbalance $2\delta$ should be proportional to the charge gap $\Delta_{\rm CO}$ and to the effective correlations $V/W$.
In the corresponding plot, Fig.~\ref{fig:2delta}, we illustrate that \NO\ nicely follows the linear dependence between $2\delta$ obtained from vibrational spectroscopy and $\Delta_{\rm CO}$ extracted from the transport measurements.
This
observation suggests that the CO state is mainly driven
by Coulomb repulsion that occurs from charge imbalance
between neighboring sites in the organic stack.  Here the anion potential seems to play only a minor role.
Interestingly, \NO\ does not follow the general relation between $T_{\rm CO}$ and $2\delta$ \cite{pustogow2016electronic}; as a matter is fact, the charge order transition occurs at about twice the temperature as expected.

\subsection{Anion order}

At elevated temperatures the anions of Fabre salts are randomly oriented; due to entropy reasons the non-centrosymmetric anions prefer to become oriented upon cooling.
The increasing charge disproportionation on the TMTTF stacks below $T_{\rm CO}$ and their coupling to the anions, however, is not sufficient to completely order the anions.
From our dc resistivity measurements
(Fig.~\ref{fig:transport}) we see that the AO transition takes place only at $T_{\rm AO}=47$~K, in coarse agreement with previous results \cite{coulon1982new,moret1983structural}.
Although this is a structural  transition, we do not observe a hysteresis in the resistivity data,
in good agreement with previous results \cite{pouget1996structural,pouget2012structural}.
The small size of the NO$_3^-$ anion might be one reason why the phase transition inclines towards second order. This also implies a weak coupling to the TMTTF molecules and minor influence on the electronic properties of \NO, as seen in the insignificant change of the resistivity below $T_{\rm AO}=47$~K.

The particular shape and the small size of the NO$_3^-$ anions explain, why the $T_{\rm AO}$ is
far below the transition temperature observed in (TMTTF)$_2$ReO$_4$ (157K) and (TMTTF)$_2$ClO$_4$ (73.4K) which contain much large anions. The other Fabre salt with tetrahedral anion, (TMTTF)$_2$BF$_4$, seems to make an exception, with $T_{\rm AO} = 41.5$~K. For tetrahedral anions, sulfur-ligand interactions govern the charge distribution in the AO state, while for \NO\ -CH$_3\cdots$O interactions are more important; we will come back to this in the subsequent Section~\ref{sec:tetramerization}. In the case of the planar NO$_3^-$ anions, the three-dimensional rotation of the entire molecule might be frozen at much higher temperatures --~for instance at $T_{\rm CO}$~-- while rotation within the pane continues down to $T_{\rm AO}$.

It is interesting to note that both branches of the $\nu_{28}({\rm A}_g)$ vibrational mode reveals a slight upshift at $T_{\rm AO}$, as seen in Fig.~\ref{fig:v28}(b). This could indicate a minute reduction in the overall charge transfer. Or it indicates a general stiffening of the vibrational pattern due to interaction with the anion, which move closer to the TMTTF molecules.
\begin{figure}
\centering
\includegraphics[width=0.7 \columnwidth,clip]{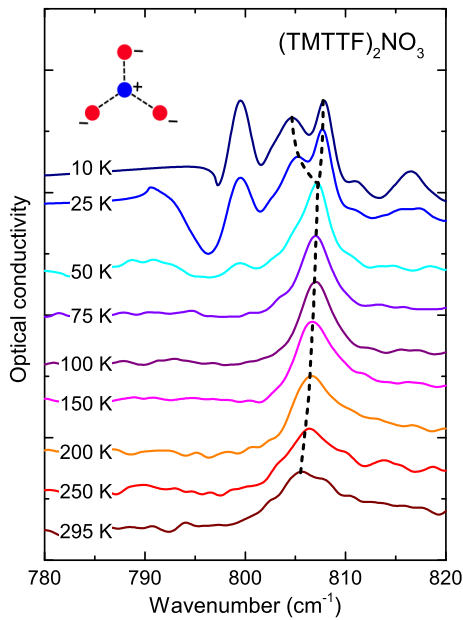}
\caption{Temperature evolution of the optical conductivity in the region of the fundamental $\nu_2$ mode of NO$_3^-$ anion measured for $E \parallel c$-axis. The curves are displaced for clarity. The dashed black line tracks the resonance frequency of the $\nu_2$ molecular vibration upon cooling and visualizes the splitting below $T_{\rm AO}=50$~K.}
\label{fig:NO3mode}
\end{figure}

A deeper insight into the anion ordering is gained by looking at the vibrations of the NO$_3^-$ anions.
the out-of-plane bending mode $\nu_2$ shows up rather prominently around $810-820$~\cm\ \cite{stewart1964infrared,mathieu1950raman}. From the  temperature dependence of this fundamental mode plotted in Fig.~\ref{fig:NO3mode}, we observe a clear splitting below $T_{\rm AO} = 50$~K.
This could simply be explained by a distortion of the NO$_3^-$ anions at the
transition resulting in breaking of the triangular symmetry.
But we can also imagine different amounts of charge on the anions
in their ordered state; in other words, the anion order causes some charge order in the NO$_3^-$ chain. This is a result of the alternating coupling to the TMTTF molecules, leading to a tetramerization there.

\subsection{Tetramerization}
\label{sec:tetramerization}
\begin{figure}[b]
\centering
\includegraphics[width=0.9 \columnwidth,clip]{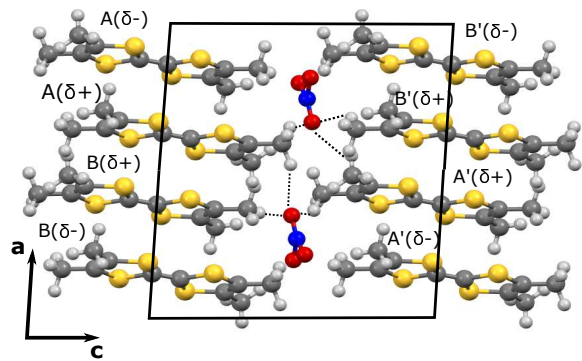}
\caption{Sketch of the \NO\ tetramer in the charge and anion ordered state at low temperatures.
Here the dimers A and A$^{\prime}$, as well as B and B$^{\prime}$ are linked via inversion symmetry.
The dashed lines indicate the shortest -CH$_3\cdots$O distances. ($\delta+$) corresponds to the charge rich sites and $(\delta-)$ to the charge poor ones; forming a 0110 pattern along the $a$-direction.}
\label{fig:tetramerization}
\end{figure}

The (TMTTF)$_2X$ stoichiometry already imposes a dimerization onto the TMTTF stacks, as depicted in Fig.~\ref{fig:structure}(a). The AO with a wavevector ($\frac{1}{2},0,0$) leads to a tetramerization
in the \NO\ salt, i.e.\ a doubling of the unit cell along the $a$-axis [Fig.~\ref{fig:structure}(b)].
Importantly, this is not just a structural order due the alternating orientation of the anions
but also a shift of the anions within the methyl-group cavity. Furthermore it affects the link towards the S atom
and rearranges the charges \cite{pustogow2016electronic,le2001temperature}.
Due to the small size and minor shift of the NO$_3^-$ anions, this structural alternation is quite weak compared to other (TMTTF)$_2X$ salts, nevertheless it is a crucial phenomenon even in the present case.
For the sister compound (TMTSF)$_2$NO$_3$ --~where sulfur atoms are substituted by Se~-- the AO transition is very similar and the wavevector is identical.
It was shown for this compound  \cite{emge1984novel} that the -CH$_3\cdots$O interactions are important;  the anion reorientation below $T_{\rm AO}$ and the shift toward selected methyl groups cause a change of the charge pattern, where donors with shorter -CH$_3\cdots$O distances are more positively charged (charge rich), while donors with smaller -CH$_3\cdots$O distances are charge poor \cite{barrans1999low,alemany2014electronic,Guster_2020}. We do expect a similar behavior for \NO:
the uniform charge distribution at elevated temperatures becomes alternated below $T_{\rm CO}$ presumably by the nearest-neighbor Coulomb repulsion $V$. In a second step, this 1010 alternation will change to a 0110 pattern in the AO state, as depicted in Fig.~\ref{fig:tetramerization}.

This corresponds exactly to the charge disproportionation within the NO$_3^-$ anions, inferred from Fig.~\ref{fig:NO3mode}: below $T_{\rm AO}$ the fundamental vibrational modes splits because every second anion becomes charge rich while the other are depleted.

\begin{figure}
\centering
\includegraphics[width=0.8 \columnwidth,clip]{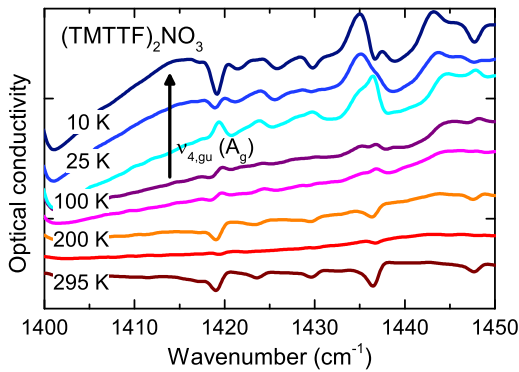}
\caption{Temperature dependence of the mid-infrared spectra of \NO\ probed with $E\parallel a$. Around 1410 to 1420~\cm\ the $\nu_{4,gu}({\rm A}_g)$ mode becomes activated by the tetramerization in  the AO state.}
\label{fig:v4}
\end{figure}

An unambiguous proof of the tetramerization along the stacks is given in Fig.~\ref{fig:v4},
where the appearance of the $\nu_{4}({\rm A}_g)$ vibration of the TMTTF molecule is
observed as $T<T_{\rm AO}$. This fully symmetric mode is typically seen in Raman spectroscopy but
infrared silent \cite{dressel2012comprehensive}, however, an out-of-phase coupling can make them
infrared active. In the present case, AO activates the dimeric in-phase vibration by out-of-phase combination of two dimers within the tetramer; the $v_{4,\rm{ gu}}$ mode couples to infrared light
via the a dipole between the dimers \cite{pustogow2016electronic}.
In Fig.~\ref{fig:v4} we see this feature at around 1410~\cm\ appearing in the $E\parallel a$ spectra below 50~K. Compare to other Fabre salts with non-centrosymmetric anions, the intensity of this mode is relatively small; but this is consistent with the fact that tetramerization in \NO\ is rather weak.
With growing anion size $d({\rm NO}_3) < d({\rm BF}_4) < d({\rm ClO}_4) < d({\rm ReO}_4)$
the tetramerization is more pronounced and the intensity of the $\nu_{4,gu}$ mode increases.
This supports the assumption that the 0110 charge arrangement in the AO state is mainly due to the anionic potential and anion arrangement; while the CO state is governed by nearest-neighbor Coulomb repulsion.

In Fig.~\ref{fig:transport} we have demonstrated that the anion-ordering transition affects the transport properties as a slight kink at $T_{\rm AO}$, but does not lead to drastic step with a hysteretic behavior. The activation energy remains unchanged, only some offset is observed;
in other words, the physical state above $T_{\rm AO}$ is very similar to the low-temperature phase.
The tetramerization on the TMTTF stack is most prominent in ESR properties \cite{coulon1982new,moret1983structural,majer2020charge} where spin singlets are formed below $T_{\rm AO}$
leading to an exponential decay of the spin susceptibility as presented in Fig.~\ref{fig:ESR2}. From the corresponding fit we obtain a spin
gap $\Delta_\sigma \approx$ 48 K in full accord with previous observations. It is interesting to note that the phase transition occurs much more gradual, compared to (TMTTF)$_2$BF$_4$ or (TMTTF)$_2$ReO$_4$ \cite{Salameh2011}, for instance.

\section{Conclusions}

Detailed transport, ESR and infrared studies on \NO\ were performed to explore the charge-ordered and anion-ordered states in this salt. From the different methods we independently determine the charge order transition at $T_{\rm CO} \approx 250$~K with a charge imbalance of $2\delta = 0.2e$.
Our findings suggest that the charge disproportionation in the CO state is mainly governed by intersite Coulomb repulsion and solely determined by the TMTTF stacks.
The universal dependence between the charge imbalance $2\delta$ and size of the CO gap $\Delta_{\rm CO}$ also holds for \NO.
In the AO state the NO$_3^-$ anions order in an alternating fashion. Due to the -CH$_3\cdots$O interaction  charge pattern changes towards 0110 below $T_{\rm AO}=50$~K. This not only causes a step in the dc resistivity, but most prominent the formation of spin singlets, leading to an exponential  drop in the spin susceptibility.

\section{Acknowledgements}
We acknowledge the support by the Deutsche Forschungsgemeinschaft and appreciate helpful discussions with A. Pustogow.

\bibliography{references_new}

\end{document}